\documentclass{article}
\usepackage{latexsym}
\usepackage{multicol}

\begin{document}
\title{\LARGE \bf A remark on the BA model of scale-free networks}
\author{{\large\bf Shinji Tanimoto\footnote{tanimoto@cc.kochi-wu.ac.jp}}\\
Department of Mathematics,
Kochi Joshi University,\\
Kochi 780-8515, Japan. \\
}
\date{}
\maketitle
\begin{abstract}
The degree distributions of many real world networks follow power-laws whose
exponents tend to fall between two and three. Within the framework of 
the Barab\'asi-Albert model (BA model), we explain this empirical observation by a simple fact.
To that end we propose a modified BA model with one parameter that
serves as a regulatory factor for the growth rate of added links in scale-free networks.
The regulatory factor has something to do with the obvious fact that one link has two nodes.
The modified model also allows to connect nodes by newly added links that do not necessarily 
emanate from new nodes.
Another related model using the master equation is also given, from which the same power-law 
degree distribution can be derived.
\end{abstract}
\vspace{0.3cm}
\begin{multicols}{2}
\begin{center}{\large \bf 1. Introduction}
\end{center}
\indent
\indent
In recent years, it has been revealed that many real networks obey the power-law degree 
distribution: 
\begin{eqnarray*}
p(k) \propto k^{- \gamma}.
\end{eqnarray*}
Here $p(k)$ is the probability that 
a randomly chosen node (or vertex) has $k$ links (or edges), and the exponent
$\gamma$ is a constant, usually 
\[
2 \le \gamma \le 3.
\]
Such networks are called
{\it scale-free} in [1]. Some examples of scale-free networks and 
their exponents are exhibited in the following table ([3]). 
\begin{center}
\begin{tabular}{|c|c|}
\hline
Network & $\gamma$ \\
\hline
\hline
WWW (indegree) & 2.1 \\
\hline
WWW (outdegree) & 2.72 \\
\hline
Internet (domain) & 2.1-2.2 \\
\hline
Internet (router) & 2.48 \\
\hline
Movie actors & 2.3 \\
\hline
Phone-call & 2.1 \\
\hline
\end{tabular}
\end{center}
\indent
\indent
There exist also a few networks whose degree distributions follow power-laws with exponents $\gamma < 2$
or $\gamma > 3$.
Traditionally, networks were thought of as random graphs, which were studied by 
Erd{\H o}s and R\'enyi in the 1960s. For those the degree distribution is a binomial or Poisson 
distribution, which is peaked about the average and decaying exponentially away from it. \\
\indent
In order to explain the above empirical fact of real networks, Barab\'asi and Albert
proposed a model (BA model) in [1]. Their model could deduce the degree distribution:
$p(k) \propto k^{-3}$, {\it i.e.}, $\gamma = 3$. \\
\indent
Now we are able to find many other network models
by which power-law degree distributions with a variety of exponents 
can be deduced (see [4, 5] or references therein). 
However, it seems that the BA model even now provides a constructive model of complex networks for 
interested researchers and students. \\
\indent
The objective of this paper is to produce scale-free networks with exponents between two and three
based on the BA model. 
For that purpose we propose a modified BA model having one parameter. The parameter
regulates the growth rate of added links in scale-free networks
and it has something to do with the obvious fact that one link has two nodes. 
Another related model using the master equation ([5]) is also given, from which the same power-law 
degree distribution can be derived. \\ 
\\
\begin{center}{\large \bf 2. A modified BA model}
\end{center}
\indent
\indent
In this section, reviewing the BA model ([1]) in more detail, we will propose a modification of
the model. The modified BA model is involved with a parameter and it coincides with 
the BA model in a particular case. We derive the condition for the parameter and
describe how it affects our model. \\
\indent
The BA model is constructed by two ingredients.
First, a network grows by continuously admitting new nodes to it and
new nodes connect to nodes already
present in the network. Thus it expands by acquiring more nodes and links over time. 
Second, the manner in which a new node connects to 
existing nodes is not completely random. The mechanism of connection is based on so-called
`preferential attachment'. This means that the probability that a new node connects to an existing node
is proportional to its degree. \\
\indent
The proposed model is as follows.
We start with a network having a small number of nodes, $m_0$, and some links. 
Notice that the value $m_0$ and the number of the original 
links are not important, since we eventually take the limit with respect to time. 
New nodes having some new links constantly enter the network. 
Whenever a new node enters the network, we advance 
the time-step by one, and $m$ new links are added to the network in each time-step, 
including the links emanating from the new node. Here $m$ is a positive integer. \\
\indent
In our model, two nodes 
which already exist in the network are allowed to connect each other.
This possibility
was not considered in the BA model ([1, 2]). Connections to
existing nodes are always executed by preferential attachment.  
Preferential attachment means that the probability, ${\mathcal P}$,
that any node connects to a node $v$ is proportional to the connectivity $k$ of that node $v$, 
so that
\begin{eqnarray} 
{\mathcal P}(k) = \frac {k}{\sum_j {k_j}}. 
\end{eqnarray} 
\indent
After $t$ time-steps, we obtain a network with $t +m_0$ nodes and $mt$ links plus the original
links. Because of the mechanism of preferential attachment, a node that happens to acquire
more links than others will continue to increase its connectivity; 
a `rich-get-richer' phenomenon. \\
\indent
Thus the rate at which a node $v_i$ with $k_i$ links acquires more links is given by
\begin{eqnarray} 
\frac{dk_i}{dt}= g{\mathcal P}(k_i),
\end{eqnarray} 
where $g$ is a proportionality constant, which is to be determined later. 
The denominator in (1), which we denote by 
\[
{\mathcal L}(t)=\sum_j {k_j},
\]
is the total of all degrees in the network at time-step $t$. In view of (1), 
taking the sums over $i$ in both sides of (2), we have 
\begin{eqnarray} 
\frac{d{\mathcal L}(t)}{dt}= g.
\end{eqnarray} 
\indent
Since $m$ links are added to the network in each time-step and each link is
counted twice in counting degrees, the total degree ${\mathcal L}(t)$ at time-step $t$ satisfies
\begin{eqnarray}
{\mathcal L}(t) \approx 2mt.
\end{eqnarray}
\indent 
Therefore, in view of (3) and (4), the growth rate $g$ of the total degree is not equal to 
$m$ as in [1, 2],
but approximately equal to $2m$. The growth rate will be equal to $m$, 
when one completely neglects the degrees of new nodes. On the other hand,
it is equal to $2m$, when one takes a full account of them. So it appears that taking $g = m$
underestimates $g$. \\
\indent
Considering the contribution of new nodes to
the growth rate $g$ to some extent, we introduce a parameter $\beta$ 
into the BA model by $g = \beta m$. It is reasonable to 
assume that $\beta$ satisfies $1 \le \beta \le 2$ from the above argument.
The parameter serves as a regulatory factor for the growing network. \\
\indent
Thus equation (2), together with (4) and $g =\beta m$, can be written as
\begin{eqnarray}
\frac{dk_i}{dt} = \beta m \frac{k_i}{2mt} = \beta \frac{k_i}{2t}.
\end{eqnarray}
For the value of $\beta$ we need the corresponding initial condition imposed upon a new node $v_i$, 
which enters the network at time-step $t_i$.
This condition is the only equation that prescribes the new node, while (3) and (5) are supposed to be 
equations for those nodes that are already part of the network. \\
\indent
From (3) we see that $g=\beta m$ is the amount of degree's increase 
for the part of the existing network within one time-step. On the other hand, 
$2m$ is the total amount of degree's increase. Thus the initial condition is given by
\begin{eqnarray}
k_i(t_i) = (2 - \beta)m,
\end{eqnarray}
for each new node.\footnote[5]{(6) is the major correction 
to the previous versions.}  \\
\indent
Next let us examine the condition for $\beta$ in further detail from (6).
It necessarily implies $\beta < 2$ and $\beta \ge 1$. The former is obvious. For the latter,
suppose $\beta < 1$. Then it follows from (6) that the number of new links emanating from
a new node is greater than $m$, which is impossible.
Therefore, the appropriate condition for the parameter $\beta$ is
\begin{eqnarray*}
1 \le \beta < 2.
\end{eqnarray*}
Moreover, in the case of $1 < \beta < 2$, notice that condition (6)
permits two nodes already present in the network to connect each other 
by some new links. The number of such
links, among $m$ newly added ones, is equal to
\begin{eqnarray*}
m - (2 - \beta)m = (\beta - 1)m.
\end{eqnarray*}
When $\beta = 1$ as in [1, 2], the possibility is forbidden. That is, in this case
all new links are connections between only a new node and nodes of the existing network. \\
\indent 
In Section 3, under the condition $1 \le \beta <2$, 
we show that this modified BA model, (5) together with (6), generates a scale-free network
with exponent 
\[
\gamma = 1 + \frac{2}{\beta}.
\] 
So we have $2 < \gamma \le 3$ for our model. In Section 4, using a quite different approach as in [5], 
a similar power-law will be obtained by means of the master
equation based on (5), too. \\
\\
\begin{center}{\large \bf 3. Scale-free networks with exponents between 2  and 3}
\end{center}
\indent
\indent
Under the assumption $1 \le \beta  <2$,
the solution of (5) with initial condition (6) is given by  
\begin{eqnarray*}
k_i(t) = (2 - \beta)m\left( \frac{t}{t_i}\right)^{\beta/2}~~~(t \ge t_i),
\end{eqnarray*}
where $t_i$ is the time when a node $v_i$ enters the network. 
It is assumed that the time interval $s=t_{i+1}-t_i$ (for each $i$) is a constant, usually $s=1$.\\
\indent
First we verify condition (4) for the model. The total degree ${\mathcal L}(t)$ equals
\[
(2-\beta)m\left[\left( \frac{t}{t_1}\right)^{\beta/2} + 
\left( \frac{t}{t_2}\right)^{\beta/2} 
+ \cdots + \left( \frac{t}{t_i}\right)^{\beta/2} \right],
\]
where $t_i$ is the maximum time-step such that $t_i \le t$. The sum in the square brackets is
approximately equal to $2(2 - \beta)^{-1}t$ by the continuous approximation: 
\[
\int_{t_1}^{t} \left( \frac{t}{x}\right)^{\beta/2}dx.
\]
Hence we see that (4) is satisfied. \\
\indent
In order to obtain the degree distribution for this model, we could adopt the method employed in [2].
However, we will use the following alternative approach, because it is easier to follow and, moreover,
it gives the same asymptotic behavior for the degree distribution as the method in [2]. \\
\indent
For the probability $p(k)$ that a randomly chosen node $v$ has a connectivity $k$, 
it suffices to estimate the number of time-steps $t_i$'s for nodes $v_i$'s, 
which satisfy the inequalities
\begin{eqnarray}
         k-1 < k_i(t) = (2 - \beta)m\left(\frac{t}{t_i}\right)^{\beta/2} \le k,
\end{eqnarray}
or alternatively
\[
 k-1/2 < k_i(t)  \le k+1/2.
\]
Let us take (7) for the estimation and rewrite it as
\begin{eqnarray*}
         t\left[\frac{k}{(2 - \beta)m}\right]^{-2/\beta} \le t_i < 
	 t \left[\frac{k-1}{(2 - \beta)m}\right]^{-2/\beta}.
\end{eqnarray*}
In the case of $s = 1$, the number of nodes satisfying (7) is
equal to the difference of both sides of the above inequalities:
\begin{eqnarray*}
  t\{(2-\beta)m\}^{2/ \beta}\{(k-1)^{-2/\beta} - k^{-2/\beta}\}, 
\end{eqnarray*}  
or
\begin{eqnarray*}
 t\{(2-\beta)m\}^{2/ \beta}\{(1 - 1/k)^{-2/\beta}-1\}k^{-2/\beta}.
\end{eqnarray*} 
Making use of $(1-1/k)^{-2/\beta} \approx 1+2{\beta}^{-1}/k$ for large $k$,
we see that the number of nodes satisfying (7) is approximated by
\[
2\beta^{-1}t\{(2-\beta)m\}^{2/ \beta}k^{-1-2/\beta}.
\]
Since at time $t$ there exist $t + m_0$ nodes, the probability $p(k)$ is thus given by
\[
\frac{2\beta^{-1}t\{(2-\beta)m \}^{2/ \beta}}{t + m_0}k^{-1-2/\beta}.
\]
In the limit $t \to \infty$ we conclude
\[
p(k) \propto k^{-(1+2/\beta)},
\]
{\it i.e.}, $\gamma = 1 + 2/\beta$, for all large $k$.\\
\\
\begin{center}{\large \bf 4. Another related model}
\end{center}
\indent
\indent
In this section we assume $1 \le \beta \le 2$.
Related to the model discussed in Section 2, we propose another 
by means of the master equation as in [5]. Making use of (5), we show 
that it also leads us to the same power-law. \\
\indent
Let $p(k, t_i, t)$ be the probability that
a node $v_i$, entering the network at time-step $t_i$, has a degree $k$ at time $t$. 
The master equation for it is given by
\begin{eqnarray}
p(k, t_i, t+1)  =  \beta \frac{k-1}{2t}\,p(k-1, t_i, t) \nonumber \\
 + ~\left(1- \beta\frac{k}{2t}\right)p(k, t_i, t),
\end{eqnarray}
because the degree of a node with $k$ links increases by one at the rate $\beta k/2t$ from (5). \\
\indent
We define the limiting probability
\begin{eqnarray*}
p(k) = \lim_{t \to \infty}\frac{1}{t} \sum_{t_i}\,p(k, t_i, t)
\end{eqnarray*}
for deriving the degree distribution. 
It follows from (8) that the recurrence relation for this $p(k)$ is given by 
\begin{eqnarray*}
(t +1)p(k) = \beta \frac {k-1}{2t}t\,p(k-1)+\left(1- \beta \frac {k}{2t}\right)tp(k),
\end{eqnarray*}
or
\begin{eqnarray*}
 p(k) = \frac{k-1}{k + 2/{\beta}}\,p(k-1).
\end{eqnarray*}
Using well-known properties of the beta function and the gamma function, it can be written as
\begin{eqnarray*}
p(k) & = & \frac{\Gamma(k)\Gamma(1+ 2/{\beta})}{\Gamma(k+1+ 2/{\beta})}\,p(1) \\
   & = & {\rm B}(k, 1+2/{\beta})\,p(1)  \\
   & \propto & k^{-(1+2/{\beta})},
\end{eqnarray*}
for all large $k$. The last asymptotic behavior follows from Stirling's formula.\\
\begin{center}
{\large \bf References}
\end{center}
\begin{itemize}
\item[1.] A.-L. Barab\'asi and R. Albert,  
Emergence of scaling in random networks, {\it Science} {\bf 286}, 509--512, 1999.
\item[2.]A.-L. Barab\'asi, R. Albert and H. Jeong, Mean-field theory for scale-free random networks,
{\it Physica} A {\bf 272}, 173--187, 1999.
\item[3.] A.-L. Barab\'asi, Emergence of scaling in complex networks, 
{\it Handbook of Graphs and Networks}, edited by S. Bornholdt and H. G. Schuster, Wiley-VCH,
69--84, 2003.
\item[4.] S. Boccaletti, V. Latora, Y. Moreno, M. Chavez and D.-U. Hwang,
Complex networks: Structure and dynamics, {\it Physics Reports} {\bf 424}, 175--308, 2006.
\item[5.] S. N. Dorogovtsev and J. F. F. Mendes, {\it Evolution of Networks: From Biological  Nets
to the Internet and WWW}, Oxford Univ. Press, Oxford, 2003.
\end{itemize}
\end{multicols}
\end{document}